\documentclass[reprint,aip,apl,twocolumn,superscriptaddress,preprintnumbers,amsmath,amssymb,floatfix]{revtex4-1}

\usepackage{graphicx}
\usepackage[mathlines]{lineno}
\usepackage{xspace} 
\usepackage{wasysym}

\begin{document}
\title{Position and energy-resolved particle detection using phonon-mediated microwave kinetic inductance detectors}



\author{D.C.~Moore}  \email{davidm@caltech.edu} \affiliation{Division of Physics, Mathematics \& Astronomy, California Institute of Technology, Pasadena, CA 91125 }
\author{S.R.~Golwala} \affiliation{Division of Physics, Mathematics \& Astronomy, California Institute of Technology, Pasadena, CA 91125 } 
\author{B.~Bumble} \affiliation{Jet Propulsion Laboratory, California Institute of Technology, Pasadena, CA 91109 }
\author{B.~Cornell} \affiliation{Division of Physics, Mathematics \& Astronomy, California Institute of Technology, Pasadena, CA 91125 }
\author{P.K.~Day}  \affiliation{Jet Propulsion Laboratory, California Institute of Technology, Pasadena, CA 91109 }
\author{H.G.~LeDuc}  \affiliation{Jet Propulsion Laboratory, California Institute of Technology, Pasadena, CA 91109 } 
\author{J.~Zmuidzinas}  \affiliation{Division of Physics, Mathematics \& Astronomy, California Institute of Technology, Pasadena, CA 91125 } \affiliation{Jet Propulsion Laboratory, California Institute of Technology, Pasadena, CA 91109 }

\begin{abstract}
We demonstrate position and energy-resolved phonon-mediated detection of particle interactions in a silicon substrate instrumented with an array of microwave kinetic inductance detectors (MKIDs).  The relative magnitude and delay of the signal received in each sensor allows the location of the interaction to be determined with $\lesssim$1~mm resolution at 30~keV.  Using this position information, variations in the detector response with position can be removed, and an energy resolution of $\sigma_E = 0.55$~keV at 30~keV was measured.  Since MKIDs can be fabricated from a single deposited film and are naturally multiplexed in the frequency domain, this technology can be extended to provide highly-pixelized athermal phonon sensors for $\sim$1~kg scale detector elements.  Such high-resolution, massive particle detectors would be applicable to rare-event searches such as the direct detection of dark matter, neutrinoless double-beta decay, or coherent neutrino-nucleus scattering. 
\end{abstract}


\maketitle

Next generation rare-event searches such as the direct detection of dark matter require large target masses ($\sim$$10^3$~kg) with sub-keV energy resolution.  This requires increasing the mass of current solid-state, cryogenic experiments\cite{CDMSScience:2010,Angloher:2011,*Armengaud:2011} by 2 orders of magnitude, while maintaining the background-free operation of existing detectors.  Reducing the cost and time needed to fabricate and test each detector element is necessary for such large cryogenic experiments to be feasible.  

Detectors that measure both the athermal phonons and ionization created by a particle interaction have demonstrated sufficient background rejection to enable next-generation experiments\cite{Pyle:2009}.  Microwave kinetic inductance detectors (MKIDs)\cite{Day:2003,Zmuidzinas:2012} offer several advantages for providing athermal phonon sensors in large experiments relative to the transition edge sensor (TES)-based designs currently in use\cite{CDMSScience:2010,Angloher:2011,Irwin:2005}.  MKIDs can be patterned from a single deposited aluminum film, with large ($>$10~$\mu$m) features, significantly reducing fabrication time and complexity.  Since MKIDs are naturally multiplexed in the frequency domain, hundreds of sensors can be read out on a single coaxial cable, enabling a more granular phonon sensor that is expected to provide enhanced background rejection. In addition to dark matter direct detection, high-resolution, massive particle detectors are applicable to the detection of neutrinoless double-beta decay\cite{Avignone:2008} and coherent neutrino-nucleus scattering\cite{Freedman:1974}. 

Previous designs\cite{Moore:2009,Golwala:2008,Mazin:2006} attempted to absorb the incident energy in large-area collectors coupled to smaller volume, distributed MKIDs.  Although separating the absorber and sensor allowed increased sensitivity by concentrating the absorbed energy, test devices suffered from poor transmission of quasiparticles from the absorber to sensor.  Here we present a simplified design that eliminates the absorber by directly collecting the energy using large-area MKIDs.  A similar design developed independently by Swenson~et~al.\cite{Swenson:2010} has been used to demonstrate time-resolved phonon-mediated detection of high-energy interactions from cosmic rays and natural radioactivity using MKIDs.  We have previously demonstrated energy-resolved detection of 30~keV x-rays\cite{Moore:2011} using phonon-mediated MKIDs fabricated from high-resistivity nitride films\cite{Leduc:2010,*Barends:2008}, but the energy resolution was limited to $\sigma_E = 1.2$~keV by systematic variations in the reconstructed energy with interaction location.  In this work, we present results for devices fabricated from aluminum films that exhibit improved uniformity. These devices allow the location of the interaction to be determined and a position-based correction to be applied, recovering a reconstructed energy resolution within 40\% of the baseline resolution.

To calculate the expected detector resolution, we assume that the phonons uniformly illuminate the substrate surface and that incident energy is converted into quasiparticles in the inductive section of the resonators with overall efficiency $\eta_{ph}$.  In this case, the phonon lifetime, $\tau_{ph}$, is determined by the substrate geometry and fraction, $\eta_{fill}$, of the total surface area, $A_{sub}$, that is covered by MKIDs.  To resolve the rising edge of the phonon signal requires that the quality factor $Q \leq 10^4$ for resonant frequencies $f_0 \approx 3$~GHz to give a resonator response time $\tau_{res} = Q/\pi f_0 \approx 1$~$\mu$s.  For $\eta_{fill} \approx 5$\% as in existing TES-based designs\cite{Pyle:2009}, $\tau_{ph} \approx 1$~ms, while quasiparticle lifetimes, $\tau_{qp}$, in our films are measured to be 10\textendash100~$\mu$s.  Given these characteristic times, we assume $\tau_{res} \ll \tau_{qp} \ll \tau_{ph}$.  The resonator quality factor, $Q^{-1} = Q_c^{-1} + Q_i^{-1}$, depends on both the coupling quality factor, $Q_c$, due to energy losses through the coupling capacitor and the internal quality factor, $Q_i$, due to all other losses (e.g., dissipation due to quasiparticles).  For our devices, typically $Q_i > 10^6$, so $Q_c \ll Q_i$ is required for sufficient sensor bandwidth to resolve the phonon rise time.  We assume the steady-state quasiparticle number is dominated by the readout power~\cite{Visser:2012fk}, $P_{read}$, (i.e., thermal quasiparticles or those generated by stray pair-breaking radiation are negligible) so $N_{qp} = \eta_{read}P_{read}\tau_{qp}/\mathrm{\Delta}$, where $\Delta$ is the superconducting gap and $\eta_{read}$ is the efficiency for the readout power to create quasiparticles in the resonator.  While the detailed mechanism for quasiparticle generation from the sub-gap readout phonons remains unclear~\cite{Visser:2012fk}, in the most conservative case we can assume $\eta_{read} = 1$.  Finally, we assume that frequency-based readout is used with the noise dominated by the HEMT amplifier with noise temperature, $T_N$.

Given these assumptions, the expected energy resolution for an MKID-based phonon-mediated detector is\cite{Moore:2011,Zmuidzinas:2012}:
\begin{equation}
\label{eq:eres_final}
\sigma_E = \frac{\Delta}{\eta_{ph} \beta (f_0,T)} \sqrt{\frac{\eta_{read}}{\alpha p_t} \frac{A_{sub} k_b T_N}{2\pi f_0}\frac{N_0 \lambda_{pb}}{\tau_{qp} S_1(f_0,T)} }
\end{equation}
where $p_t$ is the probability for a phonon to be transmitted from the substrate to the MKID, $\alpha$ is the fraction of the total inductance of the resonators due to kinetic inductance, $\lambda_{pb}$ is the mean free path for phonons to break Cooper pairs in the MKID~\cite{Kaplan:1976,Mrzyglod:1994}, and $N_0$ is the single-spin density of states.  The ratio of the frequency to dissipation response is defined as $\beta$, while $S_1$ is a dimensionless factor of order unity describing the dissipation response at temperature, $T$\cite{Mattis:1958,Gao:thesis}.

\begin{figure}[t]
\centering
\includegraphics[width=\linewidth,keepaspectratio]{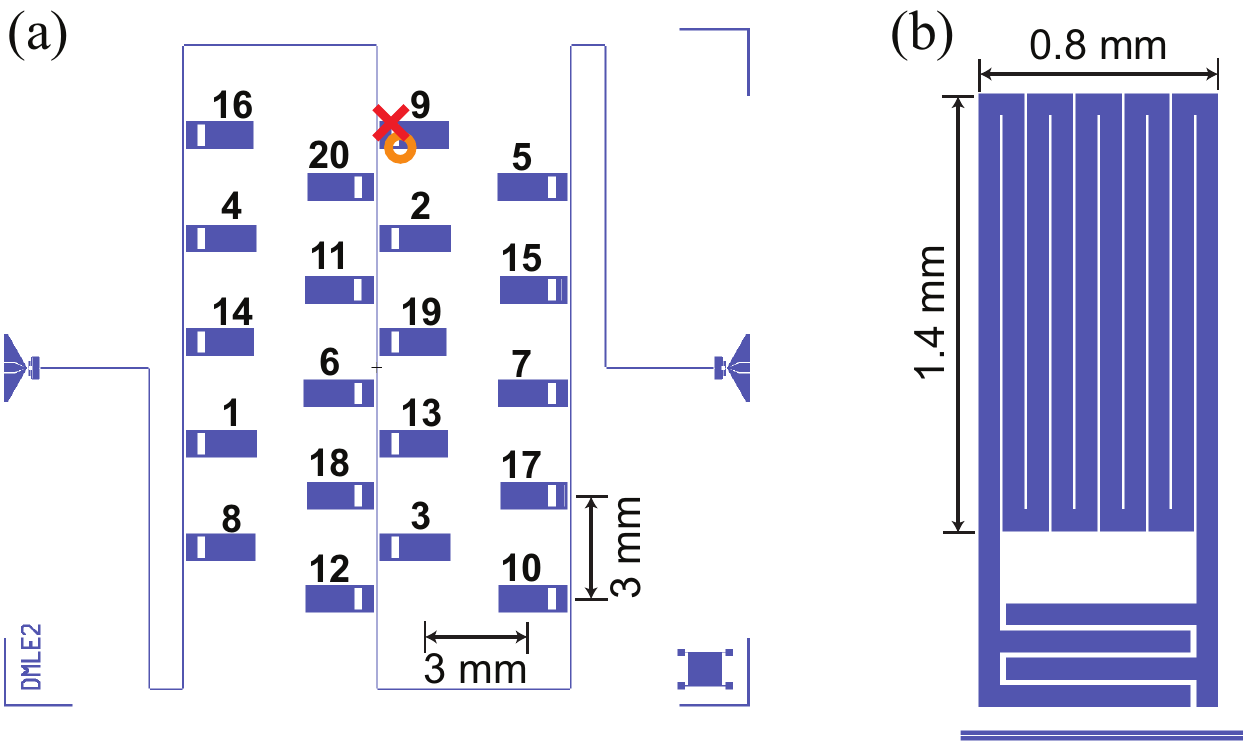}
\caption{(a)~Array layout.  The resonators are arranged in an offset grid and coupled to a coplanar strip feedline. The numbers denote the ordering of resonators by frequency, which were evenly spaced by 10~MHz so that the entire array fit within 200~MHz of bandwidth around 4.8~GHz.   The interaction locations for the event from Fig.~\ref{fig:pulse_fit} determined from the energy partitioning ($\times$) and timing delays (\Circle) are also shown.   (b)~Geometry for a single resonator.}
\label{fig:array_layout}
\end{figure}

Figure~\ref{fig:array_layout} shows the design of prototype arrays consisting of 20 lumped-element resonators\cite{Doyle:2008}, each with a 70~$\mu$m wide meandered inductor (with total area $A_{ind} = 1$~mm$^2$) connected to an interdigitated capacitor with 20~$\mu$m spacing between fingers.  Devices were fabricated on 1~mm thick, high-resistivity Si substrates ($\rho \geq 5$~k$\Omega$~cm), which were deglazed in dilute hydrofluoric acid (HF) just prior to vacuum processing.  The resonators were patterned by inductively coupled plasma etching (BCl$_3$/Cl$_2$) of a single, 25~nm thick aluminum film deposited by dc magnetron sputtering at ambient temperature.  The substrates were mounted to a testing enclosure with GE varnish at the corners of the chip, with total contact area of 8~mm$^2$.  An $^{129}$I source was mounted to illuminate the substrate face opposite the resonators, and the enclosure was cooled to 50~mK in a dilution refrigerator.


\begin{figure}[t]
\centering
\includegraphics[width=\linewidth,keepaspectratio]{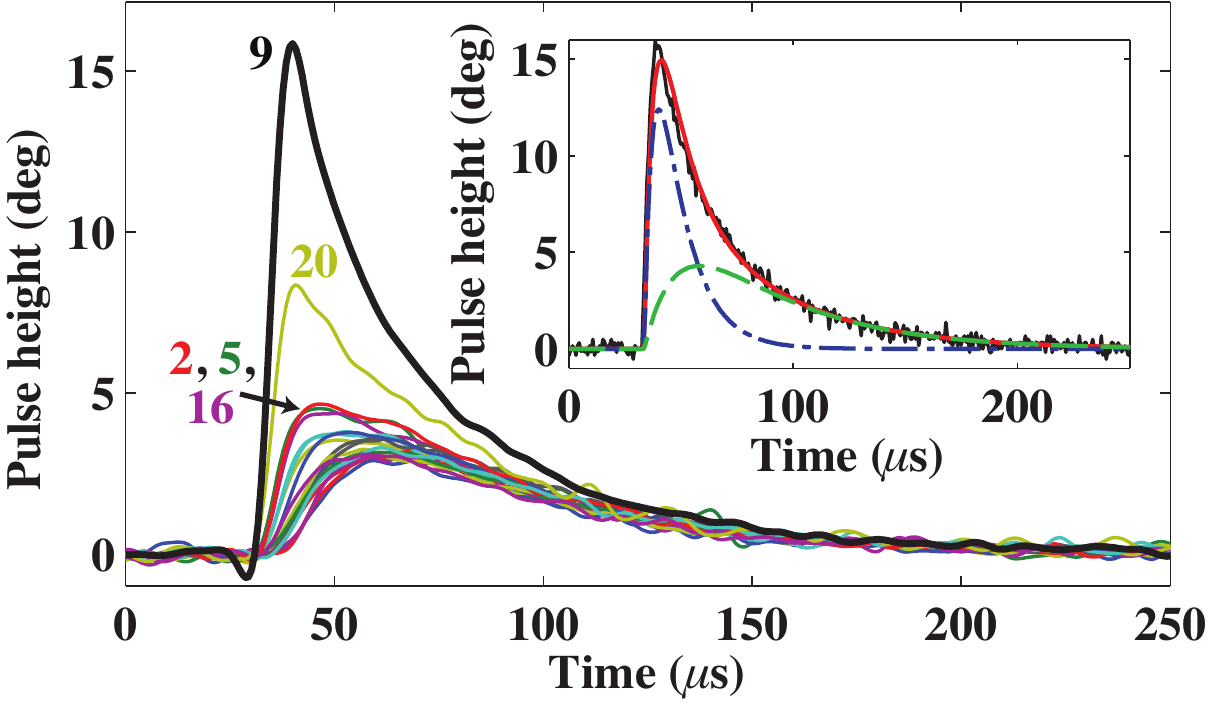}
\caption{(Color online) Pulses observed in each resonator for a 200~keV energy deposition.  The pulse heights, giving the phase shift of the signal transmitted past the resonator in degrees, have been calibrated to remove responsivity variations, and a 200~kHz low-pass filter has been applied.  The numbers denote the corresponding positions shown in Fig.~\ref{fig:array_layout} for the resonators closest to the interaction site, which have large, prompt responses. The resonators farther from the interaction have smaller, delayed responses. (inset)~Optimal filter fit (solid) to the pulse in the primary channel.  The pulse contains a prompt component (dash-dotted) that decays away with the quasiparticle lifetime as well as a delayed component (dashed) that decays away with the phonon lifetime.}
\label{fig:pulse_fit}
\end{figure}

The measured $Q_c$ varied from 5$\times$10$^3$\textendash5$\times$10$^4$, leading to a corresponding variation in the responsivity.  The relative responsivity was calibrated using $\sim$200 events from cosmic rays or natural radioactivity that deposited $\gtrsim$500~keV in the substrate.  An example 200~keV interaction is shown in Fig.~\ref{fig:pulse_fit}.  At times longer than $\sim$100~$\mu$s after the interaction, the initial position-dependent information has decayed away and the phonons have diffused sufficiently that they are uniformly distributed throughout the substrate. The phonon velocity is 6\textendash9~mm/$\mu$s, depending on polarization~\cite{Hall:1967}, indicating that the average phonon has traversed the substrate more than 20 times in this period, and all position-dependent information has been lost. By matching the amplitude of the pulses' exponential tails at long times, the relative responsivity can be determined since all resonators receive the same flux~\cite{Pyle:2009}.  After calibrating for the responsivity, we denote the ``primary channel'' as the resonator showing the largest response, and the ``opposite channels'' as the group of resonators showing the smallest response.


As shown in Fig.~\ref{fig:pulse_fit}, the pulse shape varies significantly between the primary and opposite channels.  We find that the measured pulses, $p(t)$, are well described by the sum of two components plus a noise term: $p(t) = A_{p}s_{p}(t) + A_{o}s_{o}(t) + n(t)$, where $A_{p}$ is the amplitude of the prompt component that decays with the quasiparticle lifetime, $A_{o}$ is the amplitude of the delayed component that decays with the phonon lifetime, and $n(t)$ is a noise realization.  The pulse shape for the prompt component and delayed component is given by $s_{p}$ and $s_{o}$, respectively. The standard optimal filter formalism then gives the amplitude estimates for each pulse component for the $i$th channel as:
\begin{equation}
\begin{pmatrix}
\hat{A}_{p,i} \\[6pt]
\hat{A}_{o,i}
\end{pmatrix}
 = M^{-1}
\begin{pmatrix}
\displaystyle \sum \mathrm{Re} \left(\tilde{s}_p^* \tilde{p}_i\right)J^{-1} \\[6pt]
\displaystyle \sum \mathrm{Re} \left(\tilde{s}_o^* \tilde{p}_i\right)J^{-1}
\end{pmatrix}
\end{equation}
where the sum runs over the length of the trace, $\tilde{s}_{p,o}$ and $\tilde{p}_i$ denote the Fourier transforms of $s_{p,o}$ and $p_i$, $J$ is the power spectral density of the noise and:
\begin{equation}
M = 
\begin{pmatrix}
\displaystyle \sum |\tilde{s}_p|^2J^{-1} & \displaystyle \sum \mathrm{Re} (\tilde{s}_p^* \tilde{s}_o)J^{-1}  \\[8pt]
\displaystyle \sum  \mathrm{Re} (\tilde{s}_p^* \tilde{s}_o)J^{-1} & \displaystyle \sum |\tilde{s}_o|^2J^{-1}
\end{pmatrix} 
\end{equation}

\begin{figure}[t]
\centering
\includegraphics[width=\linewidth,keepaspectratio]{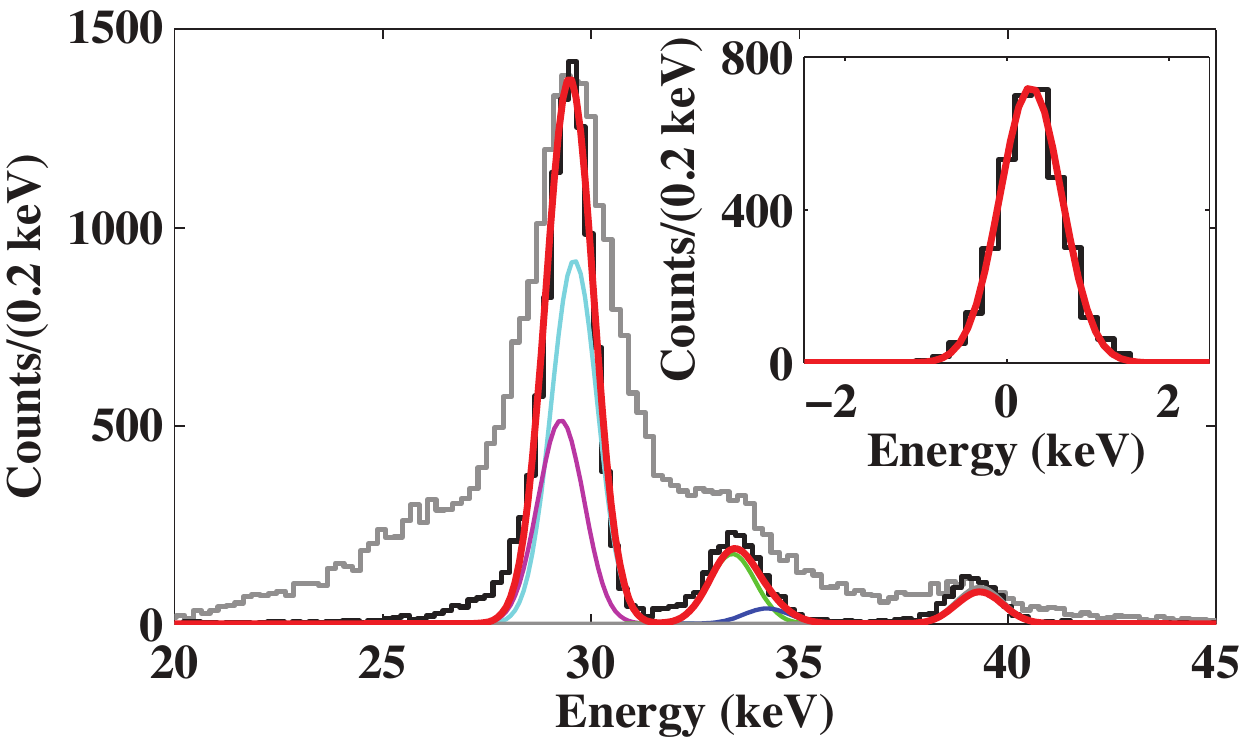}
\caption{(Color online) Observed spectrum from an $^{129}$I source.  The reconstructed energy before (light histogram) and after (dark histogram) restricting to events interacting in the central portion of the substrate and applying the position-based correction to the energy are shown.  The spectrum is fit to the observed lines at 29.5 (29.8\%), 29.8 (53.1\%), 33.6 (10.2\%), 34.4 (2.2\%), and 39.6~keV (4.6\%), where the numbers in parentheses denote the expected absorbed intensities in 1~mm of Si.  The thick line shows the fit to the total spectrum, with best fit resolution of $\sigma_E=0.55$~keV at 30~keV.  The location of the 39.6~keV peak indicates there is a $\sim$2\% non-linearity over the 30\textendash40~keV range, requiring a non-linear energy calibration similar to existing TES-based athermal phonon mediated detectors\cite{Akerib:2005zy}. (inset)~Fit to the reconstructed energy spectrum for randomly triggered noise traces giving a baseline resolution of $\sigma_E=0.38$~keV.}
\label{fig:spec_plot}
\end{figure}

The energy of the pulse is then reconstructed from the optimal filter amplitudes as $E = E_0 \sum_i (\hat{A}_{p,i} + w\hat{A}_{o,i})/r_i$, where the latter component is weighted by the ratio of the template integrals, $w = \sum s_o/\sum s_p$, and the index $i$ runs over the resonators.  Since the two pulse components measure the quasiparticle creation rate convolved with the quasiparticle decay time, $w$ gives the ratio of the total number of quasiparticles created in these components, noting that the linear convolution does not affect the ratio.  The relative responsivity of each resonator is given by $r_i$ while the overall scaling, $E_0$, is determined by calibration with x-ray lines of known energy.  Figure~\ref{fig:spec_plot} shows the reconstructed energy spectrum for phonon-mediated events from an $^{129}$I source.  

The frequency and dissipation response of the resonators with temperature was used to determine $\alpha = 0.075 \pm 0.005$ and $\Delta = 204 \pm 21$~$\mu$eV\cite{Mattis:1958,Gao:thesis}.  Comparing the response versus temperature to the response from the source indicates that $\eta_{ph} = 0.070 \pm 0.011$.  Measurements of the frequency noise confirm that it is consistent with HEMT noise for $T_N = 5$~K.  The quasiparticle lifetime, $\tau_{qp} = 12.9 \pm 1.2$~$\mu$s, was determined from the fall time of the prompt component of the primary channel pulses. Using these parameters and Eq.~\ref{eq:eres_final}, the expected resolution for this device was $\sigma_E$=(0.48~keV)$\sqrt{\eta_{read}/p_t}$, consistent with the measured baseline resolution for $\sqrt{\eta_{read}/p_t}\approx 0.8$.  The phonon lifetime was determined from the fall time of the opposite channel pulses to be $\sim$50~$\mu$s, in agreement with the 1~ms fall times seen in TES-based detectors\cite{Pyle:2009} after accounting for the differences in substrate geometry and sound speed between the devices.

\begin{figure}[t]
\centering
\includegraphics[width=\linewidth,keepaspectratio]{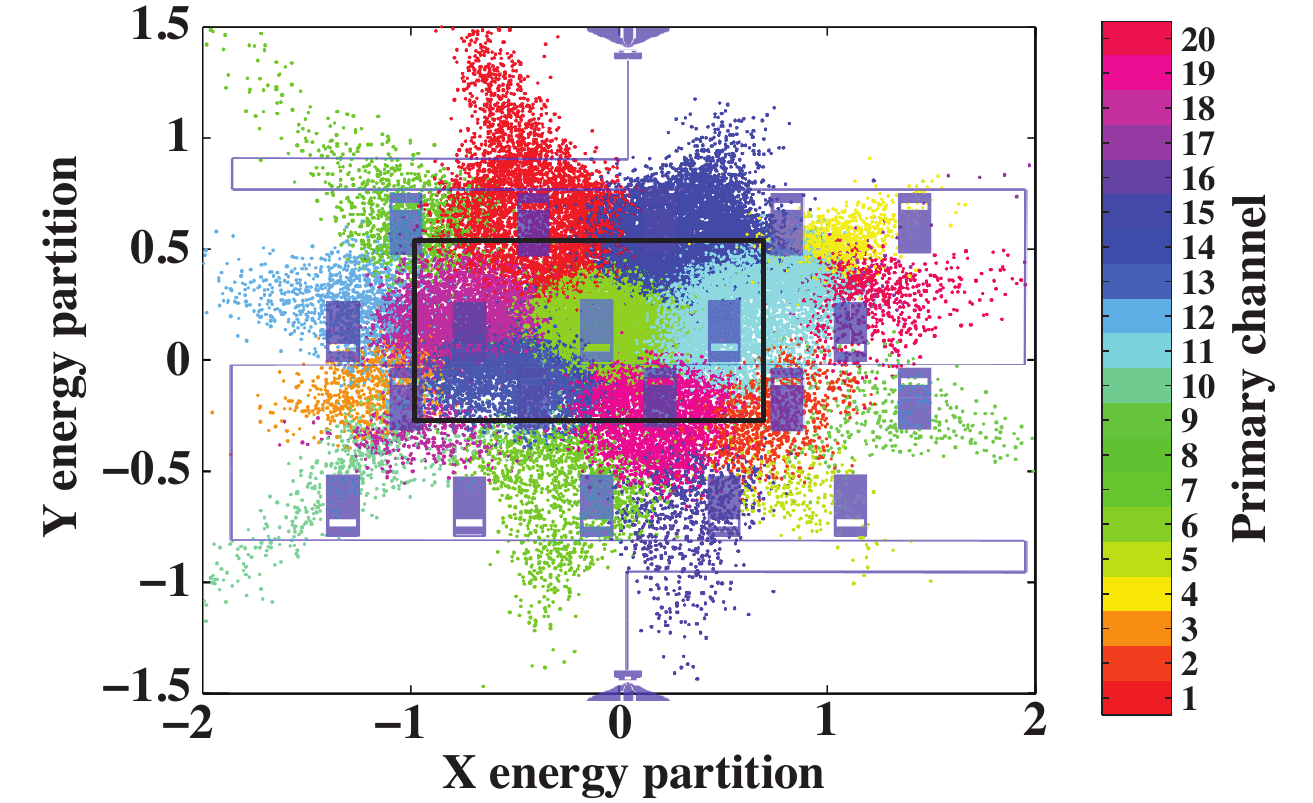}
\caption{(Color online) Reconstruction of the interaction location from the energy partition.  The coloring denotes the primary channel for each pulse.  The black lines indicate the selection of events interacting in the center of the substrate. For comparison, the device geometry from Fig.~\ref{fig:array_layout} is overlaid.}
\label{fig:part_plot}
\end{figure}

Although the baseline resolution is consistent with expectations, the measured energy resolution at 30~keV is $\sigma_E \approx 1$~keV, indicating that systematic variations in the pulse shape or amplitude with position are dominating the resolution.  In addition, non-Gaussian tails of events lie between the expected spectral peaks.  To remove poorly collected events and account for these systematic variations, a simple position estimator was constructed from the relative partitioning of energy between sensors for each event\cite{Akerib:2005zy,CDMSScience:2010}.  For $i$ resonators at locations ($x_i$,$y_i$) with energy $E_i$ detected in each resonator, the ``$X$ energy partition'' is defined as $P_x = \sum_i x_iE_i/\sum_i E_i$, and correspondingly for $Y$.  Figure~\ref{fig:part_plot} shows the reconstructed event location using this position estimator, which has higher signal-to-noise than the corresponding estimator based on timing delays at 30~keV.  Although further calibrations with collimated sources will be necessary to precisely determine the position resolution of the device, a rough estimate can be obtained from the overlap of the partition distribution for events with neighboring primary channels.  The widths of these overlap regions indicate a typical position resolution of $\sim$0.8~mm at 30~keV.

For interactions occurring near the edge of the substrate, more phonon energy can be lost to the detector housing than for interactions in the center.  Figure~\ref{fig:part_plot} shows a position-based selection of events that removes edge events with poor collection, eliminating nearly all of the events in the non-Gaussian tails between peaks.  

In addition, a position-dependent correction for variations in the reconstructed energy was applied, similar to the correction used in existing TES-based detectors\cite{Akerib:2005zy,CDMSScience:2010}.  For each event with energy partition location ($P_x$,$P_y$), the set of $n=400$ events with the closest partition location were selected.  The spectrum for these ``nearest neighbor'' events was fit to determine the location of the 29.8~keV peak, and the corrected energy for the event was determined as $E^{corr} = ($29.8~keV$/E_{29.8}^{nn})E$, where $E_{29.8}^{nn}$ is the location of the peak maximum for the event's nearest neighbors. The resulting spectrum is shown in Fig.~\ref{fig:spec_plot}.  The best-fit resolution is $\sigma_E$(30~keV$) = 0.55$~keV.  This resolution is nearly a factor of 2 better than the uncorrected resolution for all events and is within 40\% of the baseline resolution of $\sigma_E$(0~keV$) = 0.38$~keV.  

We are currently scaling this design to 0.25~kg substrates with $A_{sub} \approx 100$~cm$^2$.  Existing TES-based phonon-mediated detectors of this size have typical energy resolutions\cite{CDMSScience:2010} $\sigma_E = 1.0$~keV at 20 keV, indicating that the devices presented above can already provide comparable energy resolution.  The energy resolution of MKID-based devices can also be significantly improved.  Increasing the phonon collection efficiency to $\eta_{ph} = 30$\%, as obtained for TES-based designs with comparable metal coverage\cite{Pyle:2009}, would improve the resolution by a factor of 4.  This collection efficiency was found to scale linearly with the film thickness and may be improved by reducing phonon losses to the detector housing.  Using resonator materials with higher kinetic inductance or lower gap could improve the resolution, provided uniform resonators can be fabricated. $\tau_{qp}$ is also a factor of $\sim$100 smaller than results reported for 40~nm thick Al films at similar readout powers~\cite{Visser:2012fk}, indicating that a non-thermal quasiparticle population (possibly due, e.g., to stray light or particle interactions in the substrate) may be present in addition to readout generated quasiparticles.  If so, better shielding of the device from external radiation could improve these lifetimes significantly.  Finally, the development of lower-noise, broadband amplifiers\cite{Eom:2012}, could provide up to an additional factor of 3 improvement in resolution.   At the same time, MKIDs would provide less complex detector fabrication and higher resolution position information to enhance background rejection, simplifying the extension of these designs to the large target masses needed for future rare-event searches.

This research was carried out in part at the Jet Propulsion Laboratory (JPL), California Institute of Technology, under a contract with the National Aeronautics and Space Administration. The devices used in this work were fabricated at the JPL Microdevices Laboratory. We gratefully acknowledge support from the Gordon and Betty Moore Foundation.  This work benefited significantly from interactions with and simulation software developed by the CDMS/SuperCDMS collaborations, as well as from useful insights from B.~Mazin and O.~Noroozian.  B. Cornell has been partially supported by a NASA Space Technology Research Fellowship.

%
\bibliographystyle{apsrev4-1}

\end{document}